%% file: poster_salasj.tex
\documentclass[twoside,a4paper,11pt]{sea10}
\usepackage{graphicx}
\usepackage{hyperref}
\usepackage{movie15}
\input{mlaeff}

\begin{document}
\pagenumbering{arabic}
\pagestyle{myheadings}
\thispagestyle{empty}
{\flushleft\includegraphics[width=\textwidth,bb=58 650 590 680]{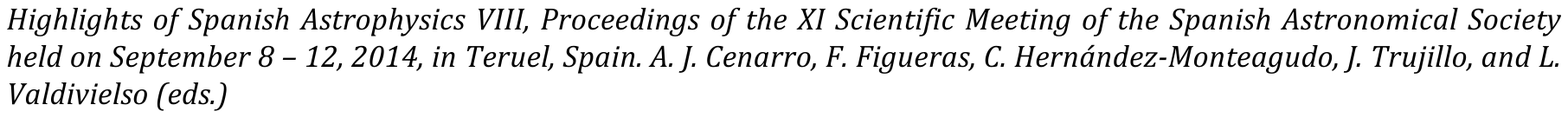}}
\vspace*{0.2cm}
\begin{flushleft}
{\bf {\LARGE
%
A photometric variability study of massive stars in Cygnus OB2
%
}\\
\vspace*{1cm}
%
J. Salas$^{1}$,
J. Ma{\'\i}z Apell{\'a}niz$^{2}$, 
and 
R. H. Barb{\'a}$^{3}$
%
}\\
\vspace*{0.5cm}
%
$^{1}$
Agrupaci{\'o}n Astron{\'o}mica de Huesca, Spain\\
$^{2}$
Centro de Astrobiolog{\'\i}a, INTA-CSIC, Spain\\
$^{3}$
Universidad de La Serena, Chile
%
\end{flushleft}
%
\markboth{
A photometric variability study of massive stars in Cygnus OB2
}{ 
%
Salas et al.
%
}
\thispagestyle{empty}
\vspace*{0.4cm}
\begin{minipage}[l]{0.09\textwidth}
\ 
\end{minipage}
\begin{minipage}[r]{0.9\textwidth}
\vspace{1cm}
\section*{Abstract}{\small
%
We have conducted a 1.5-year-long variability study of the stars in the Cygnus OB2 association, the region in the northern 
hemisphere with the highest density of optically visible massive stars. The survey was conducted using four pointings in the 
Johnson $R$ and $I$ bands with a 35~cm Meade LX200-ACF telescope equipped with a 3.2 Mpixel SBIG ST10-XME CCD camera and includes 
300+ epochs in each filter. A total of 1425 objects were observed with limiting magnitudes of 15 in $R$ and 14 in $I$. The 
photometry was calibrated using reference stars with existing $UBVJHK$ photometry. Bright stars have precisions better than
0.01~magnitudes, allowing us to detect 52 confirmed and 19 candidate variables, many of them massive stars without previous 
detections as variables. Variables are classified as eclipsing, pulsating, irregular/long period, and Be. We derive the phased
light curves for the eclipsing binaries, with periods ranging from 1.3 to 8.5 days. 
%
\normalsize}
\end{minipage}

\begin{figure}
\centerline{\includegraphics*[width=1.00\linewidth, bb=28 28 566 398]{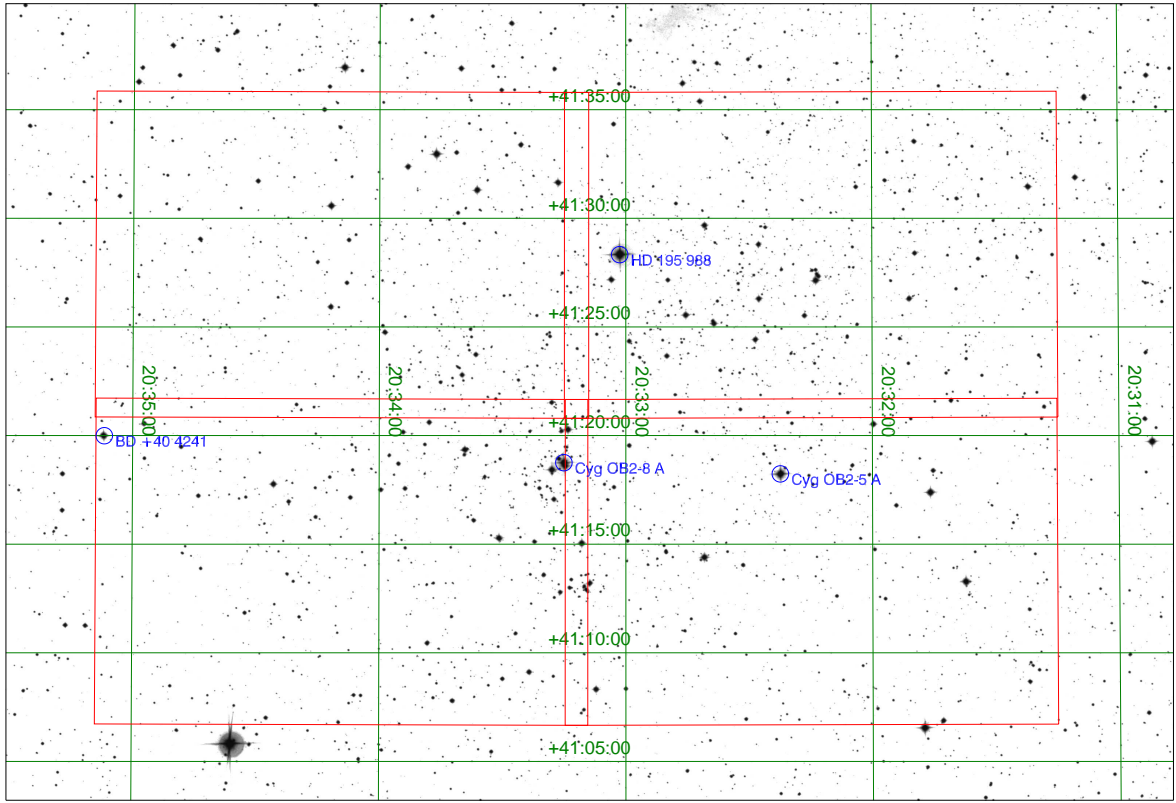}}
\caption{The four fields used in this work on a DSS2 Red image.}
\label{fig1}
\end{figure}

\begin{figure}
\centerline{\includegraphics*[width=0.95\linewidth, bb=28 28 566 566]{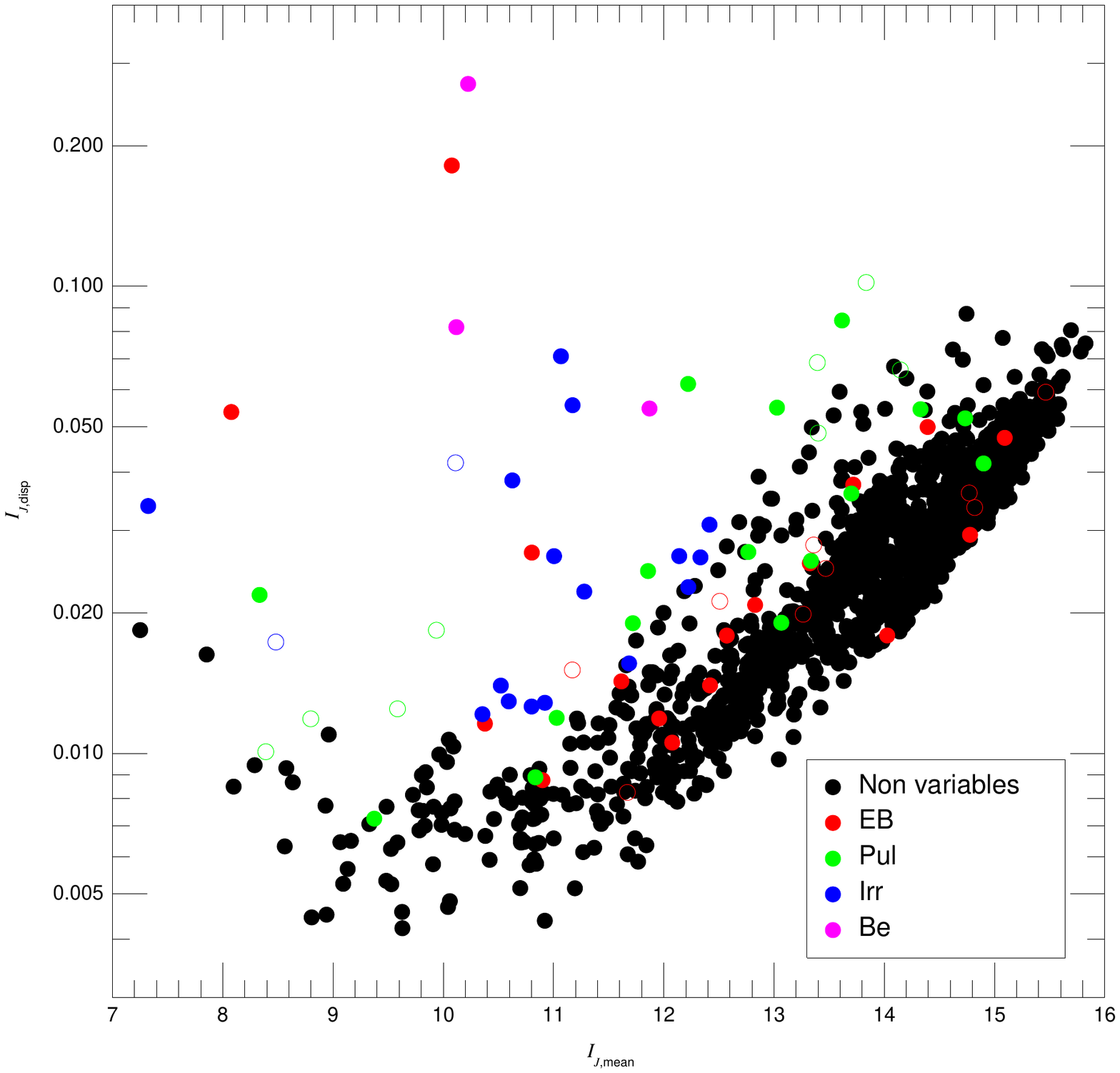}}
\caption{Dispersion - mean magnitude diagram for $I_J$. Non-filled symbols indicate candidate detections. The increase in dispersion in
the range $I_J$ = 7-8 is due to the onset on non-linearity on the CCD.}
\label{fig2}
\end{figure}   

\section{Survey description}

$\,\!$\indent We have conducted a one-and-a-half-year-long (25 June 2012 to 27 November 2013) variability study of Cygnus OB2
in the Johnson $R_J$ and $I_J$ bands (not Cousins) with Optec Inc. filters. 
We used a 35 cm Meade LX200-ACF telescope with an SBIG ST10-XME 3.2~Mpix. camera 
from a suburban location in Zaragoza, Spain, at an altitude of 260~m.
We obtained over 300 epochs per filter in
fur 22.5\arcmin$\times$15\arcmin\ fields (Figure~\ref{fig1}).
Our coverage xtends to $R_J$ = 15 and $I_J$ = 14 (S/N limited), with 1425 objects, and
there are pnly four saturated stars in the four fields: HD~195\,988, Cyg~OB2-5~A, Cyg~OB2-8~A, and BD~+40~4241 (Figure~\ref{fig1}).

\section{Why Cygnus OB2?}

$\,\!$\indent Cygnus OB2 is the best region to study O stars in the northern hemisphere for three reasons:
(a) it is young and massive, including two O3 stars (the only cases with $\delta > 0$);
(b) it has a well-populated IMF; and
it is nearby ($\sim$1.7 kpc) but it has mid-to-high extinction ($A_V$ = 4-10 mag).
Note that Cygnus OB2 is an OB association, not a cluster (and it was born that way, 
      \href{http://adsabs.harvard.edu/abs/2014MNRAS.438..639W}{Wright et al. 2014}), so it is relatively extended in the sky.
The spectral types for this work were obtained from the Galactic O-Star Catalog 
      (\href{http://adsabs.harvard.edu/abs/2004ApJS..151..103M}{Ma{\'\i}z Apell\'aniz et al. (2004)}; 
       \href{http://adsabs.harvard.edu/abs/2008RMxAC..33...56S}{Sota et al. 2008}) and the Galactic O-Star 
       Spectroscopic Survey (\href{http://adsabs.harvard.edu/abs/2011hsa6.conf..467M}{Ma{\'\i}z Apell\'aniz et al. 2011}).

\begin{table}
\caption{Confirmed and candidate eclipsing binaries.}
\centerline{$\,\!$}
\centerline{
\footnotesize
\setlength{\tabcolsep}{5pt}
\input{eb}
}
\label{tab1}
\end{table}   

\begin{table}
\caption{Confirmed and candidate pulsating variables.}
\centerline{$\,\!$}
\centerline{
\footnotesize
\setlength{\tabcolsep}{5pt}
\input{pul}
}
\label{tab2}
\end{table}   

\begin{table}
\caption{Confirmed and candidate irregular/long period variables.}
\centerline{$\,\!$}
\centerline{
\footnotesize
\setlength{\tabcolsep}{5pt}
\input{irr}
}
\label{tab3}
\end{table}   

\begin{table}
\caption{Confirmed Be stars.}
\centerline{$\,\!$}
\centerline{
\footnotesize
\setlength{\tabcolsep}{5pt}
\input{Be}
}
\label{tab4}
\end{table}   

\begin{figure}
\centerline{\includegraphics*[width=\linewidth, bb=28 28 566 566]{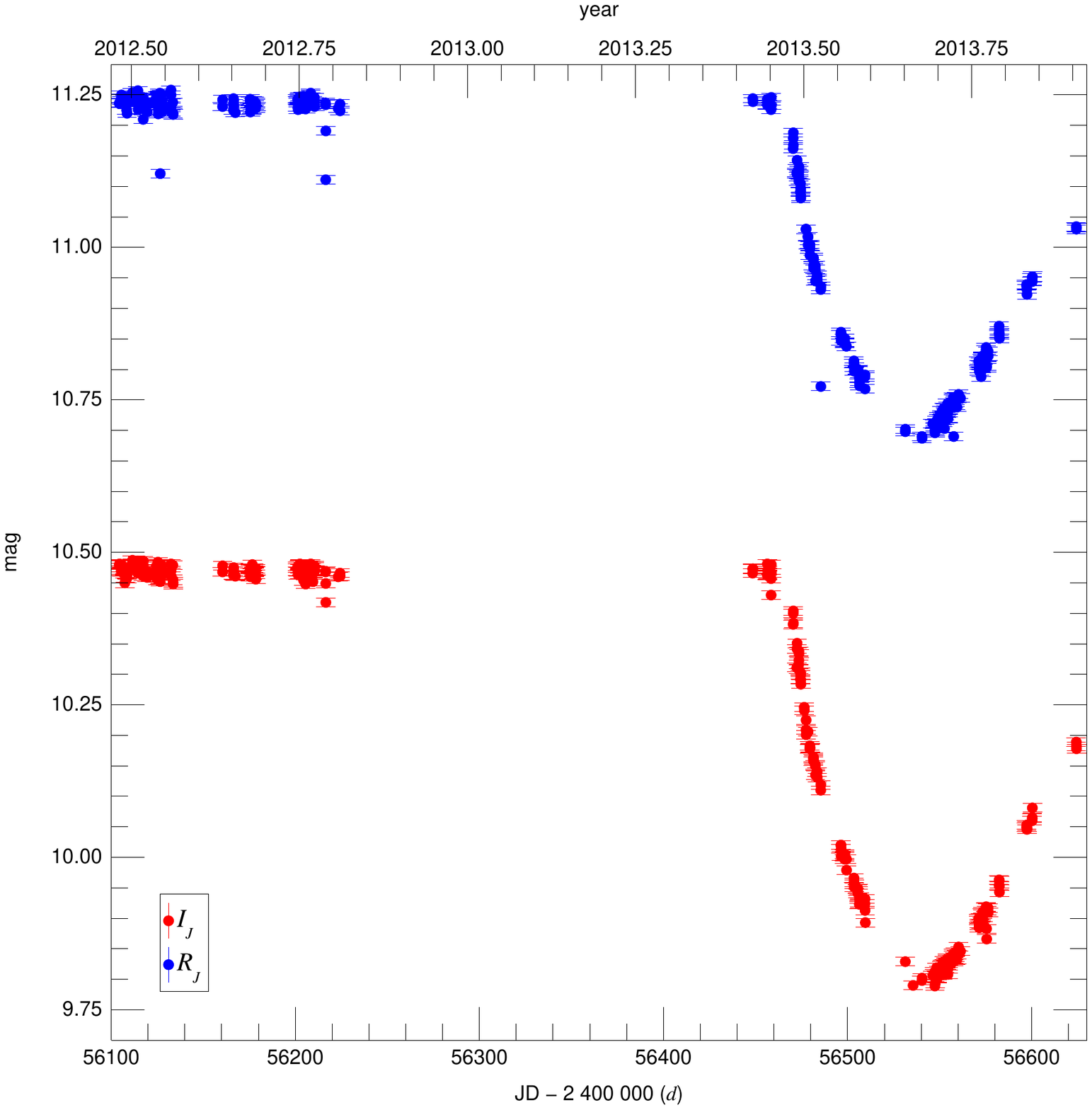}}
\caption{Light curves for Cyg OB2-4 B.}
\label{fig3}
\end{figure}

\section{Data taking and reduction}

$\,\!$\indent The observations were automated using CCD AutoPilot and MaxImDL / MaxPoint software.
The image processing was done under Pyraf with additional Python packages and scripts.
For the astrometric calibration we used \url{http://astrometry.net}.
We performed aperture photometry with four different apertures. The selection among them was based on optimal S/N and minimal neighbor contamination on 
       a case-by-case basis.
The absolute photometric calibration was performed using several reference stars by interpolating between $V_J$ from 
       \href{http://adsabs.harvard.edu/abs/1991AJ....101.1408M}{Massey \& Thompson (1991)}
       and $J$ from 2MASS using \href{http://adsabs.harvard.edu/abs/2013hsa7.conf..657M}{Ma{\'\i}z Apell\'aniz (2013)}
       and \href{http://adsabs.harvard.edu/abs/2014A&A...564A..63M}{Ma{\'\i}z Apell\'aniz et al. (2014)}.
We found a photometric dispersion of 0.01 mag or less for the calibration stars.

\section{Searching for variables}

$\,\!$\indent In order to search for variable stars, we performed a first pass analyzing the dispersion - mean magnitude diagrams (Figure~\ref{fig2}). 
In a second pass we used specific search algorithms for eclipsing binaries.
For the periodogram analyses we used both the Fourier method of \href{http://adsabs.harvard.edu/abs/1986ApJ...302..757H}{Horne \& Baliunas (1986)}
       and the information entropy method of \href{http://adsabs.harvard.edu/abs/1995ApJ...449..231C}{Cincotta et al. (1995)}. 
We considered four types of variables: 

\begin{itemize}
 \item Eclipsing.
 \item Pulsating.
 \item Irregular/long period.
 \item Be stars.
\end{itemize}

For each star we assigned a detection category from non-variable, candidate, and confirmed variable.
We compared our results with the previous studies of \href{http://adsabs.harvard.edu/abs/2011ApJS..194...27H}{Henderson et al. (2011)} 
       and \href{http://adsabs.harvard.edu/abs/2012ApJ...747...41K}{Kiminki et al. (2012)}.

\section{Results}

$\,\!$\indent We have detected:

\begin{itemize}
 \item 17 confirmed and 9 candidate eclipsing binaries (Table~\ref{tab1}).
 \item 16 confirmed and 8 candidate pulsating variables (Table~\ref{tab2}).
 \item 16 confirmed and 2 candidate irregular/long period variables (Table~\ref{tab3}).
 \item 3 spectroscopically confirmed Be variable stars (Table~\ref{tab4}).
\end{itemize}

Among the most interesting cases, we point out out that
Cyg OB2-4 B is a newly discovered Be star that underwent a brightening of 0.57 mag in $R_J$ and 0.69 mag in $I_J$ during the 
       observing period (\href{http://adsabs.harvard.edu/abs/2013ATel.5571....1S}{Salas et al. 2013}, 
       Figure~\ref{fig3}). The event was accompanied by spectral changes e.g. H$\beta$ shifted from 
       absorption to a double-peaked emission, as seen in GOSSS data.
Also, some of the eclipsing binaries have eccentric orbits (Figure~\ref{fig5}).
Finally, Cyg OB2-12 and Cyg OB2-IRS 7, two massive objects in the association, are irregular variables (Figure~\ref{fig8}).

\begin{figure}
\centerline{\includegraphics*[width=0.49\linewidth, bb=28 28 566 566]{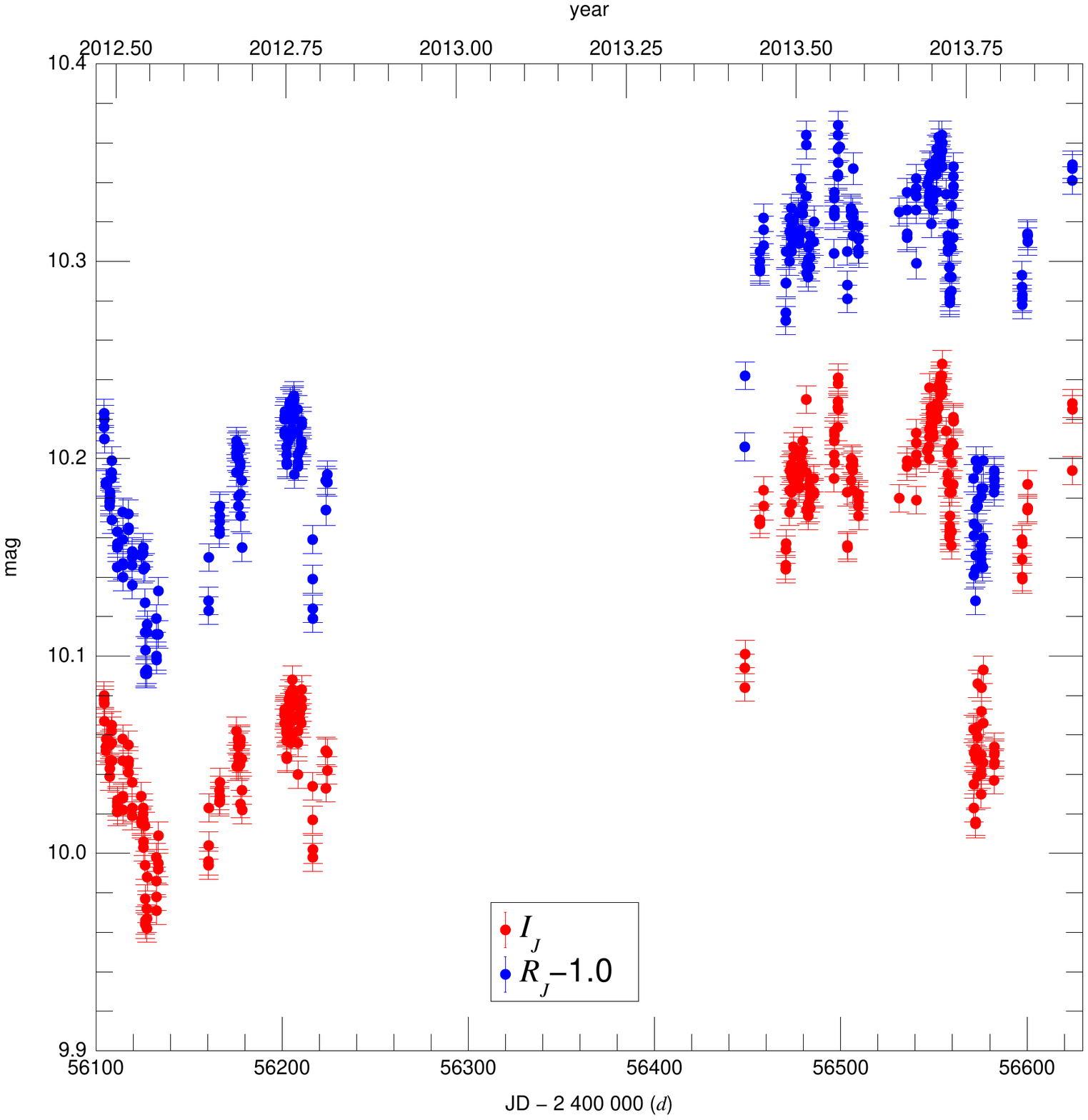} \
            \includegraphics*[width=0.49\linewidth, bb=28 28 566 566]{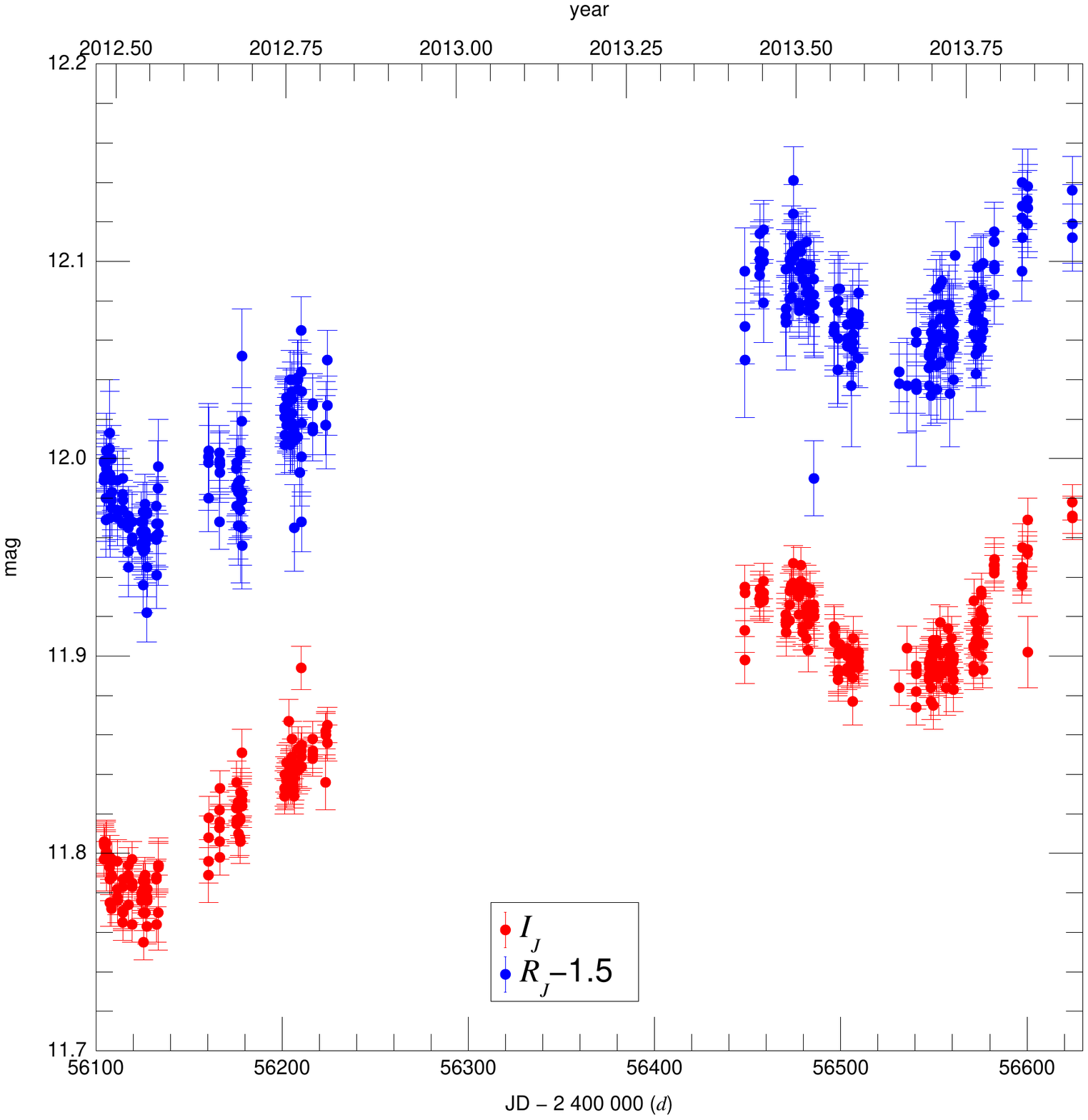}}
\caption{Light curves for two Be stars, Schulte 30 (left) and Schulte 64 (right).}
\label{fig4}
\end{figure}

\begin{figure}
\centerline{\includegraphics*[width=0.49\linewidth, bb=28 28 566 566]{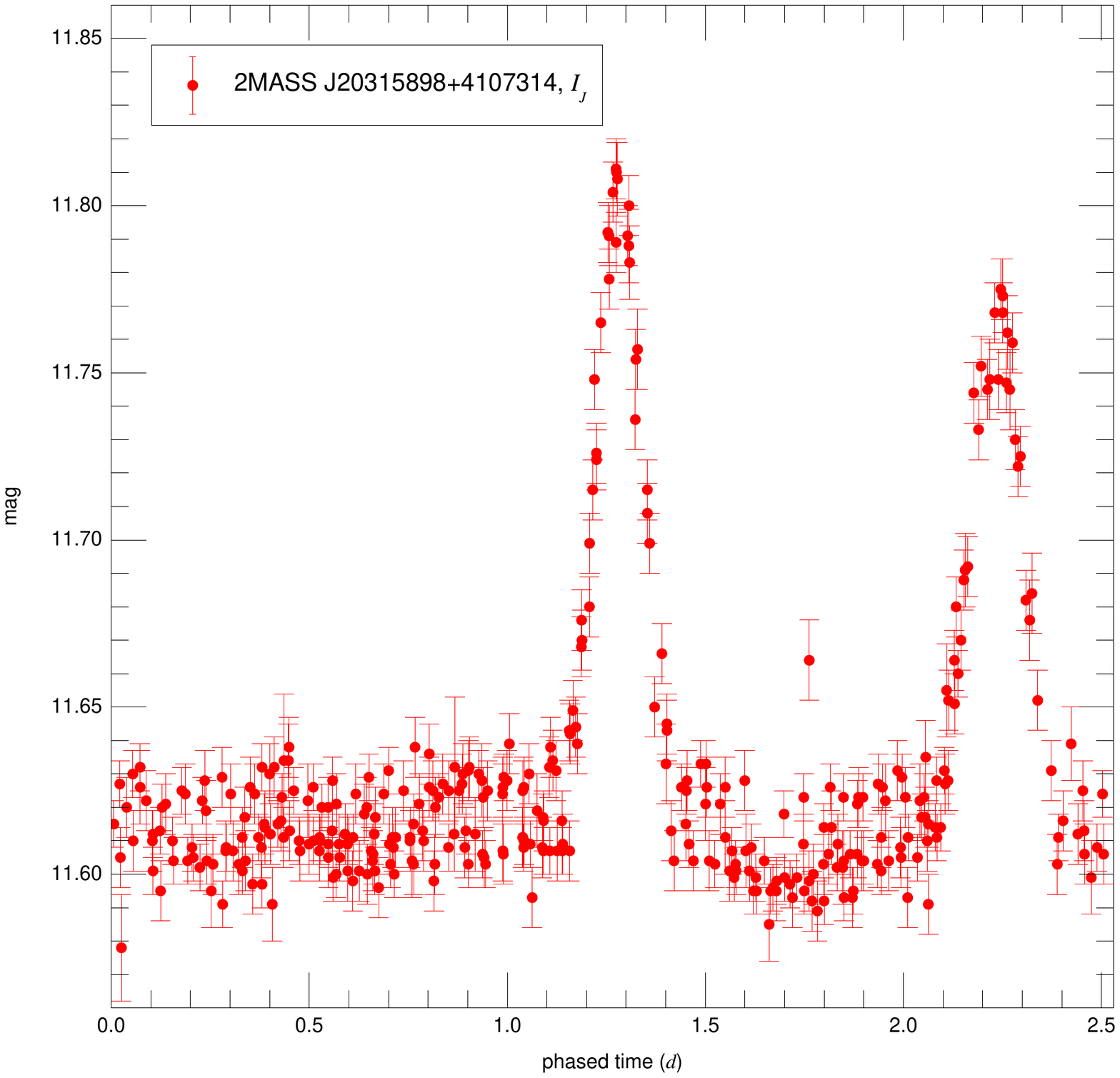} \
            \includegraphics*[width=0.49\linewidth, bb=28 28 566 566]{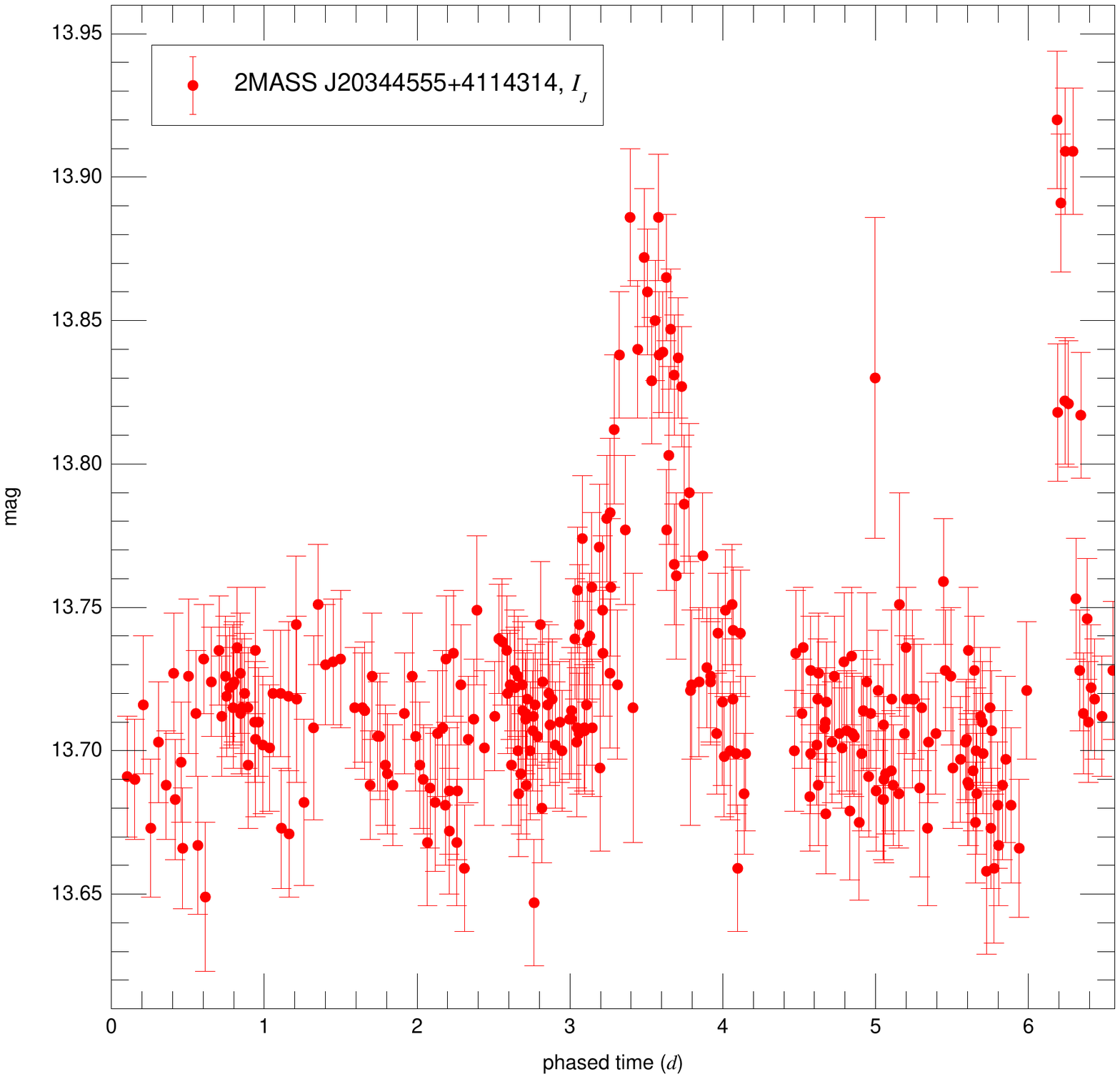}}
\centerline{\includegraphics*[width=0.49\linewidth, bb=28 28 566 566]{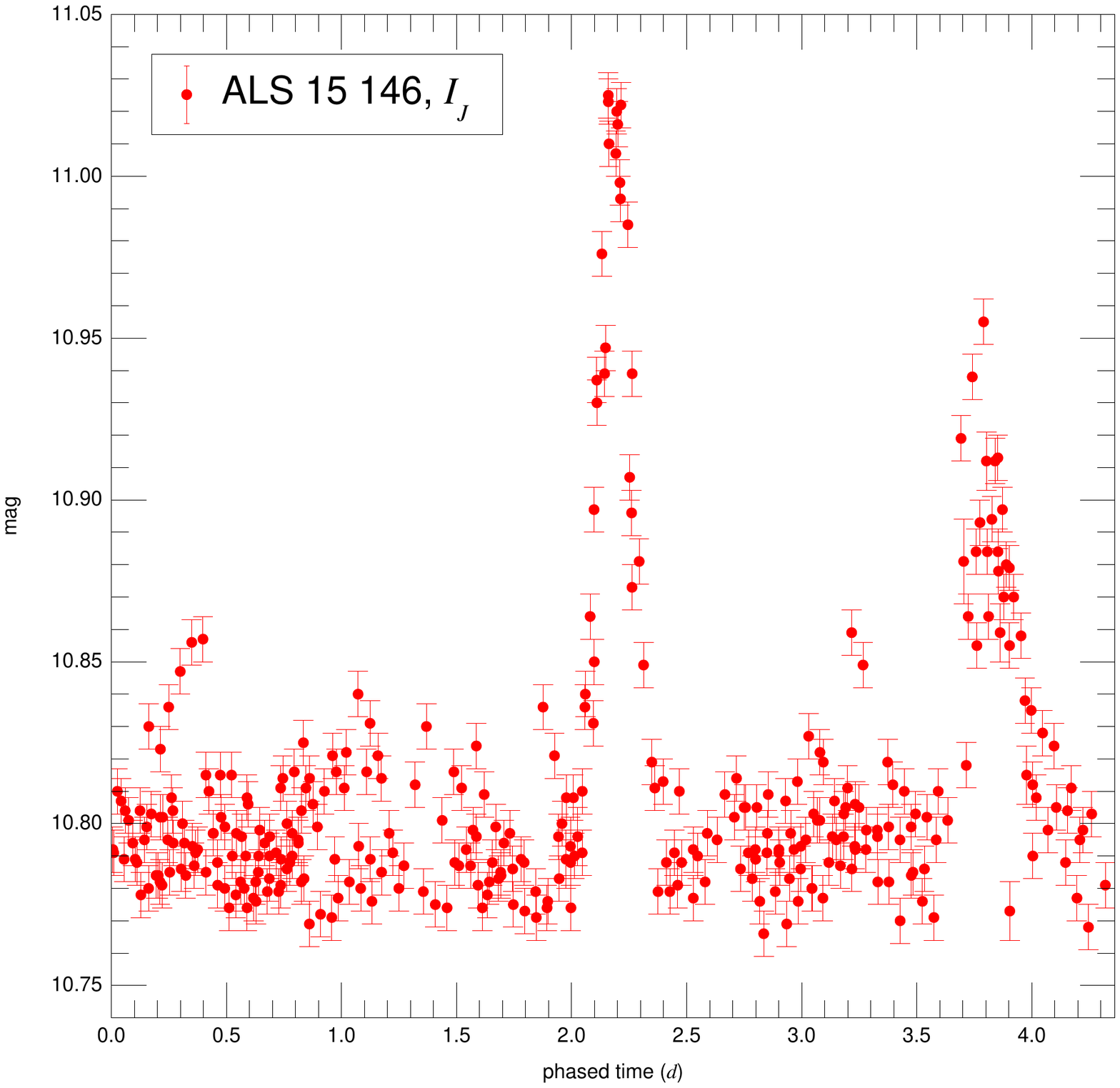} \
            \includegraphics*[width=0.49\linewidth, bb=28 28 566 566]{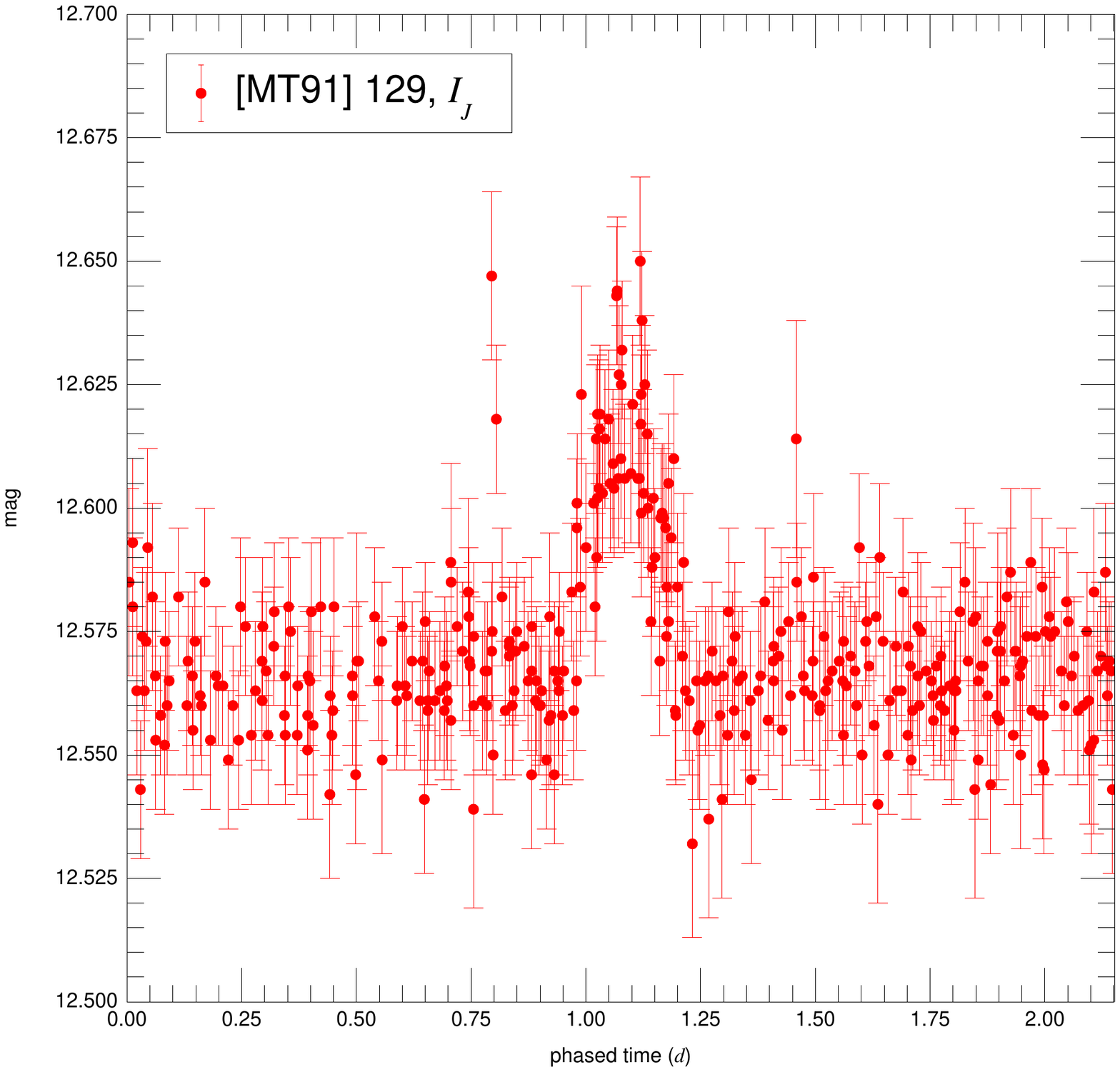}}
\caption{Phased light curves for four eclipsing binaries. Note that the first three orbits are eccentric.}
\label{fig5}
\end{figure}

\begin{figure}
\centerline{\includegraphics*[width=0.49\linewidth, bb=28 28 566 566]{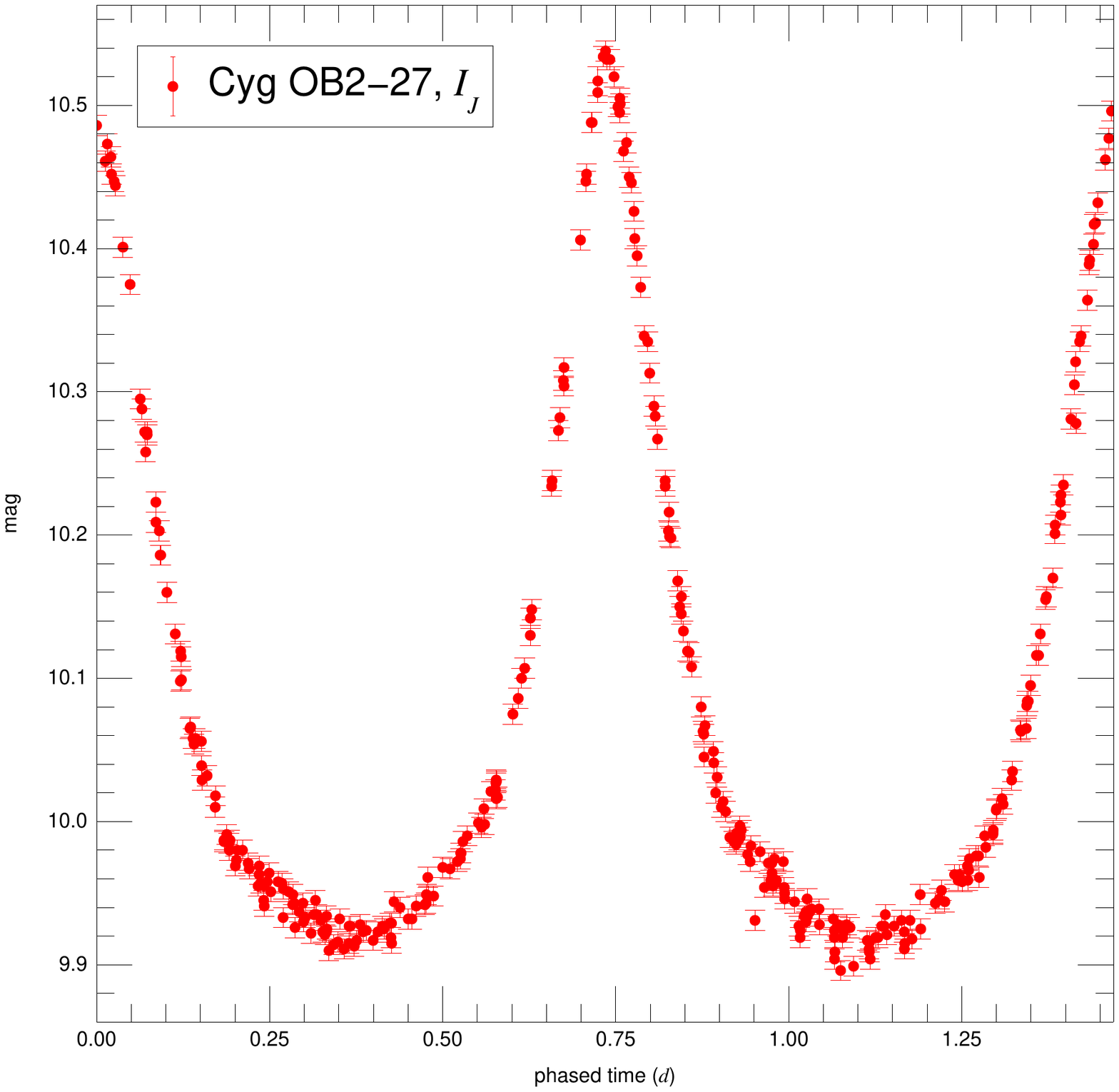} \
            \includegraphics*[width=0.49\linewidth, bb=28 28 566 566]{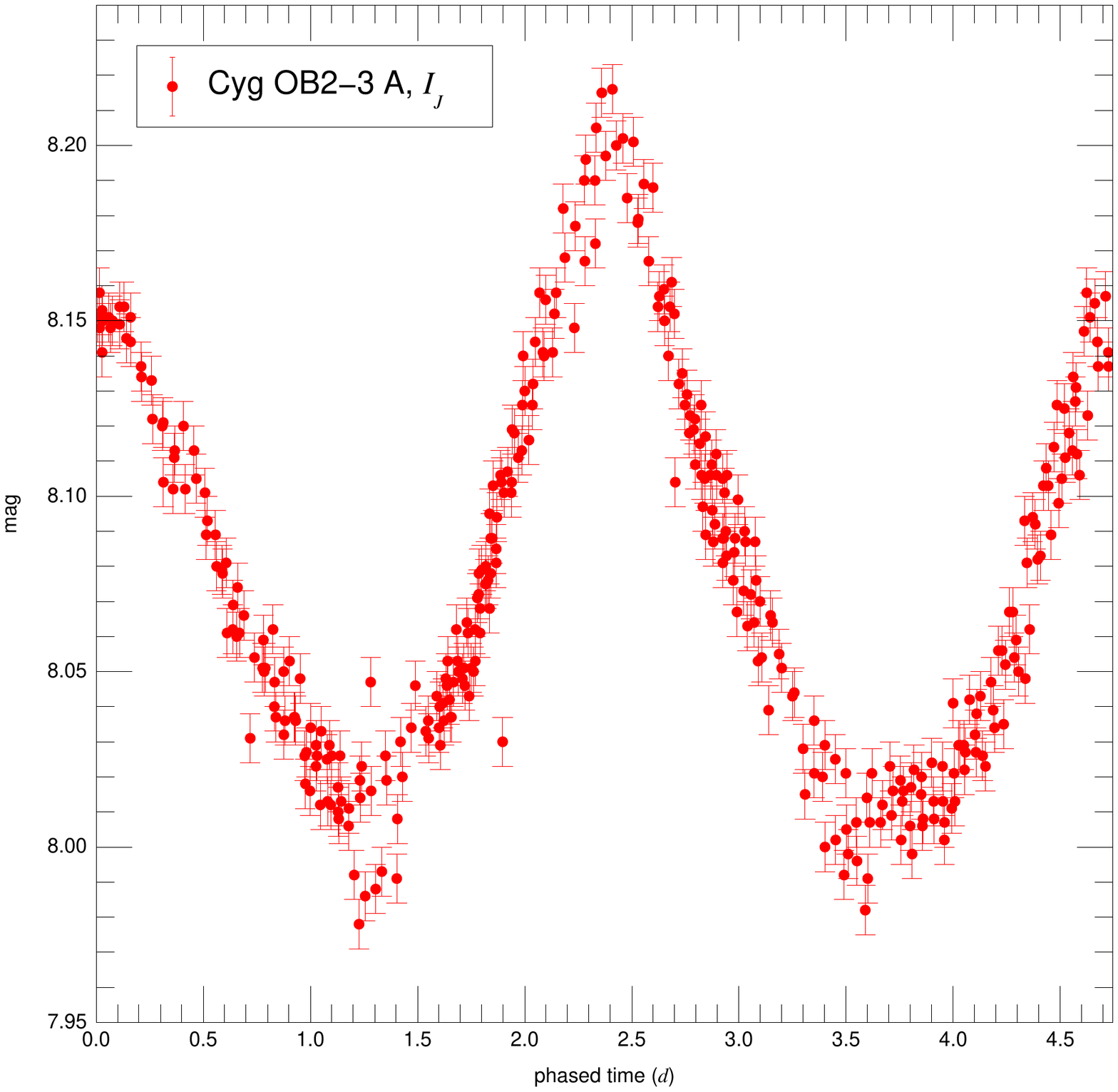}}
\centerline{\includegraphics*[width=0.49\linewidth, bb=28 28 566 566]{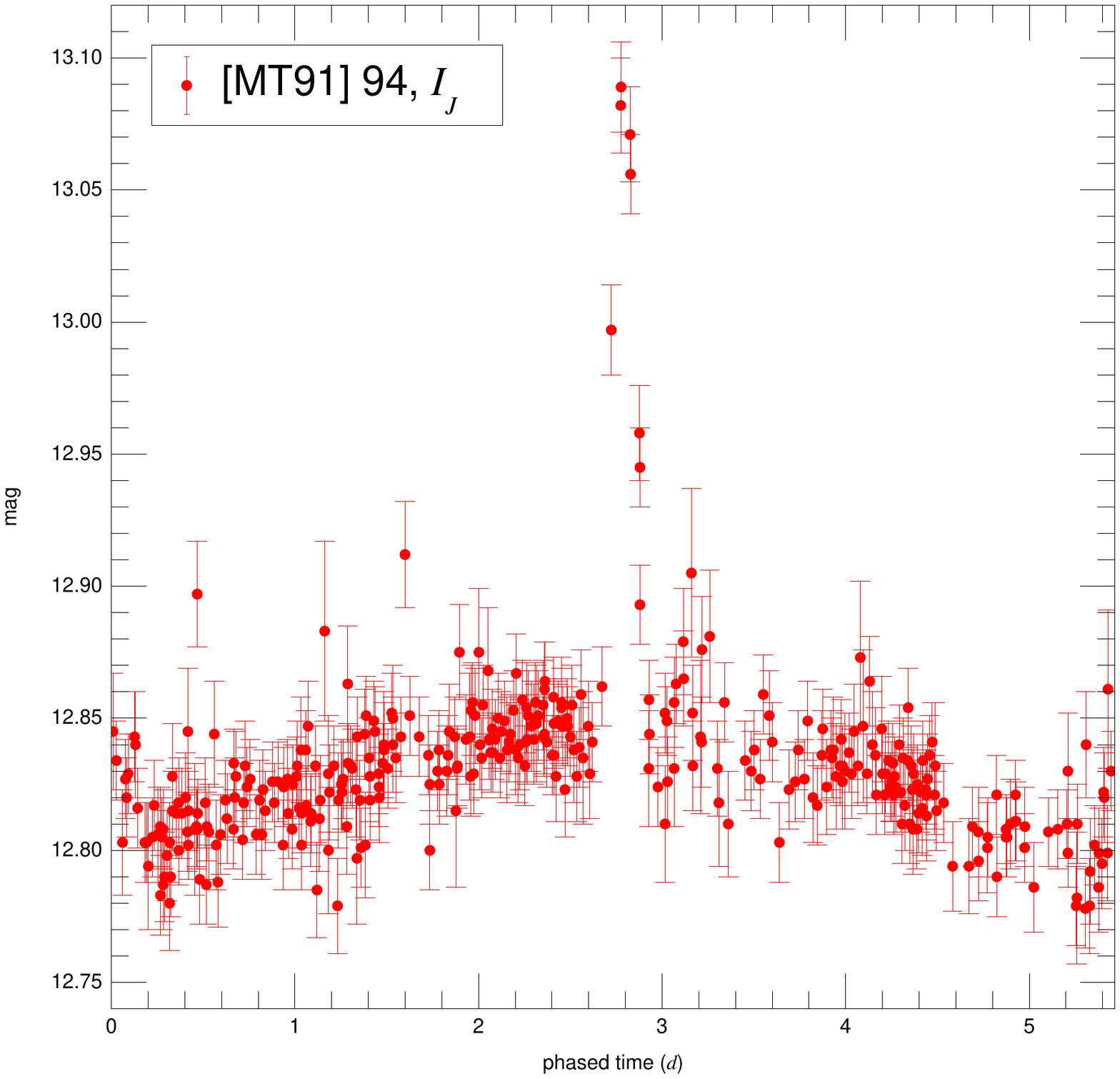} \
            \includegraphics*[width=0.49\linewidth, bb=28 28 566 566]{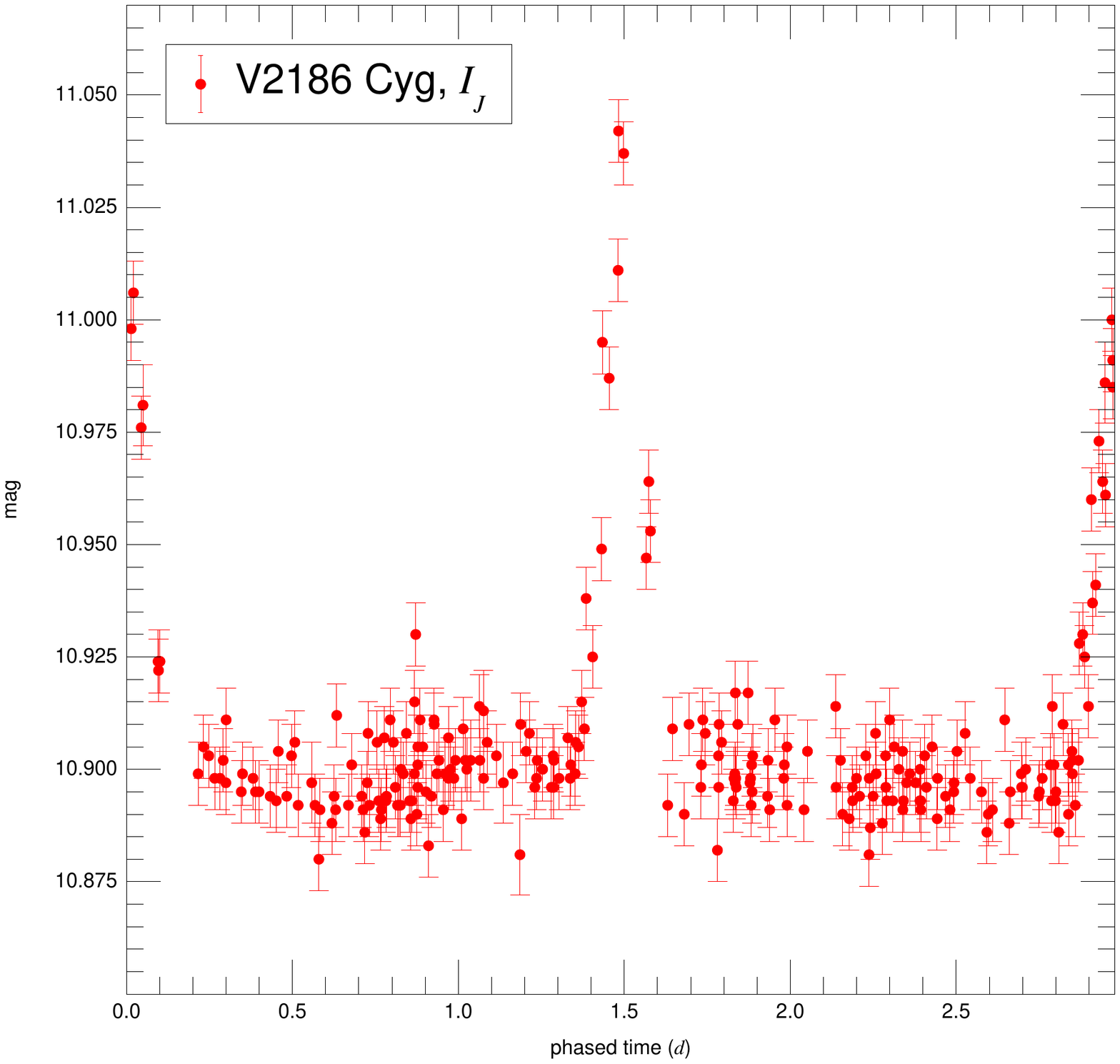}}
\caption{Phased light curves for four eclipsing binaries.}
\label{fig6}
\end{figure}

\begin{figure}
\centerline{\includegraphics*[width=0.49\linewidth, bb=28 28 566 566]{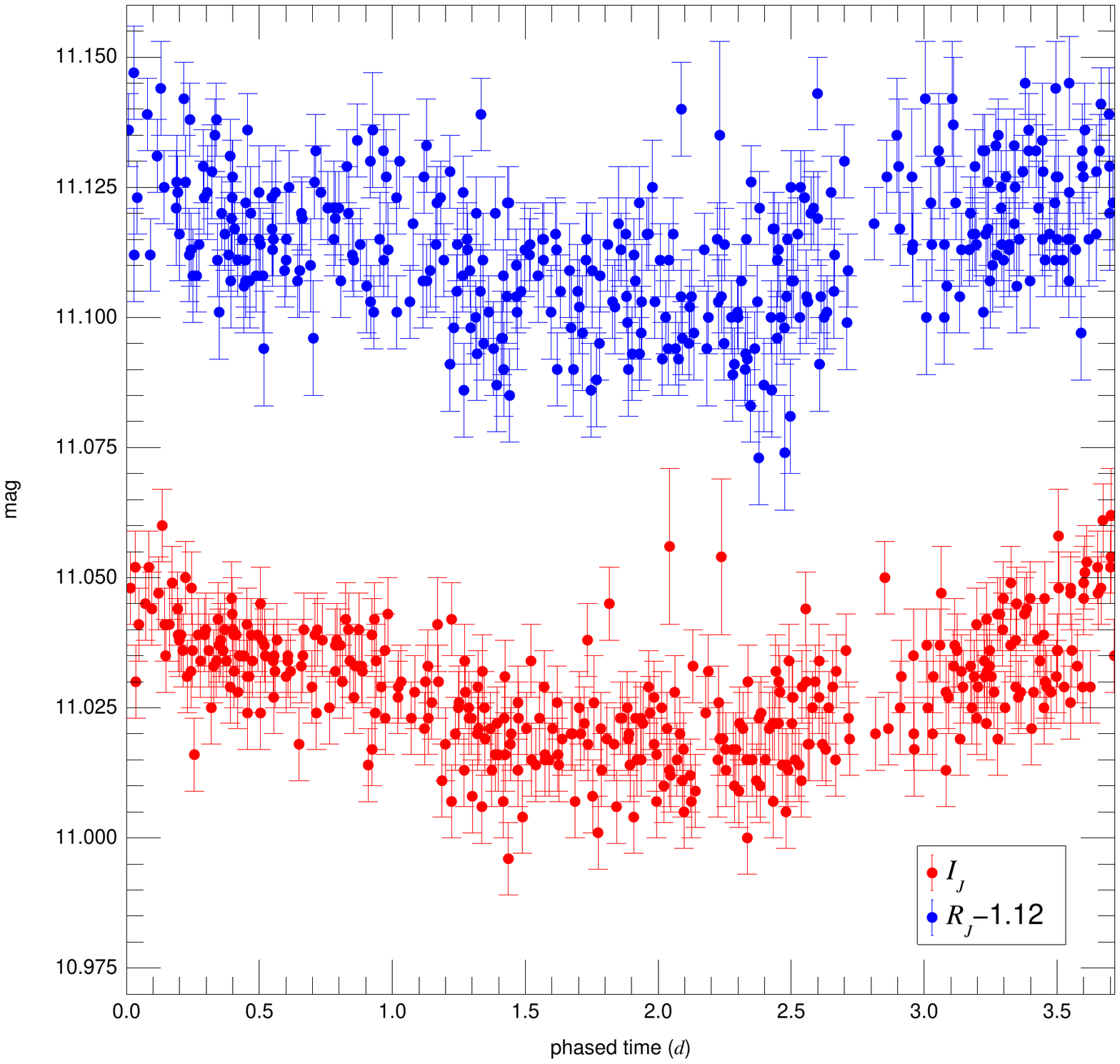} \
            \includegraphics*[width=0.49\linewidth, bb=28 28 566 566]{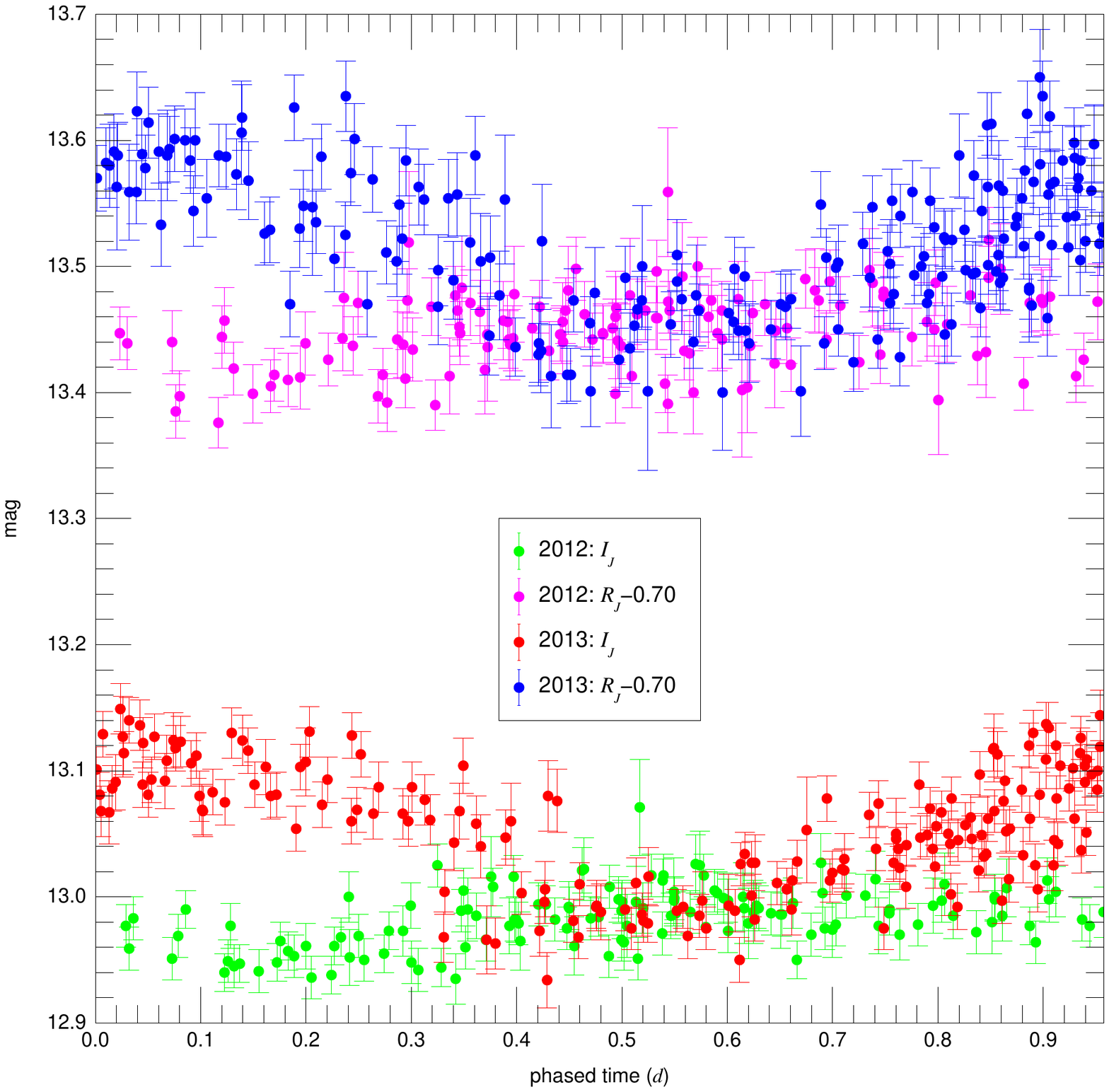}}
\caption{Phased light curves for two pulsating stars, Cyg OB2-A30 (left) and 2MASS J20334299+4130005 (right). Note how for the second
case the pulsations appear only for the 2013 data.}
\label{fig7}
\end{figure}

\begin{figure}
\centerline{\includegraphics*[width=0.49\linewidth, bb=28 28 566 566]{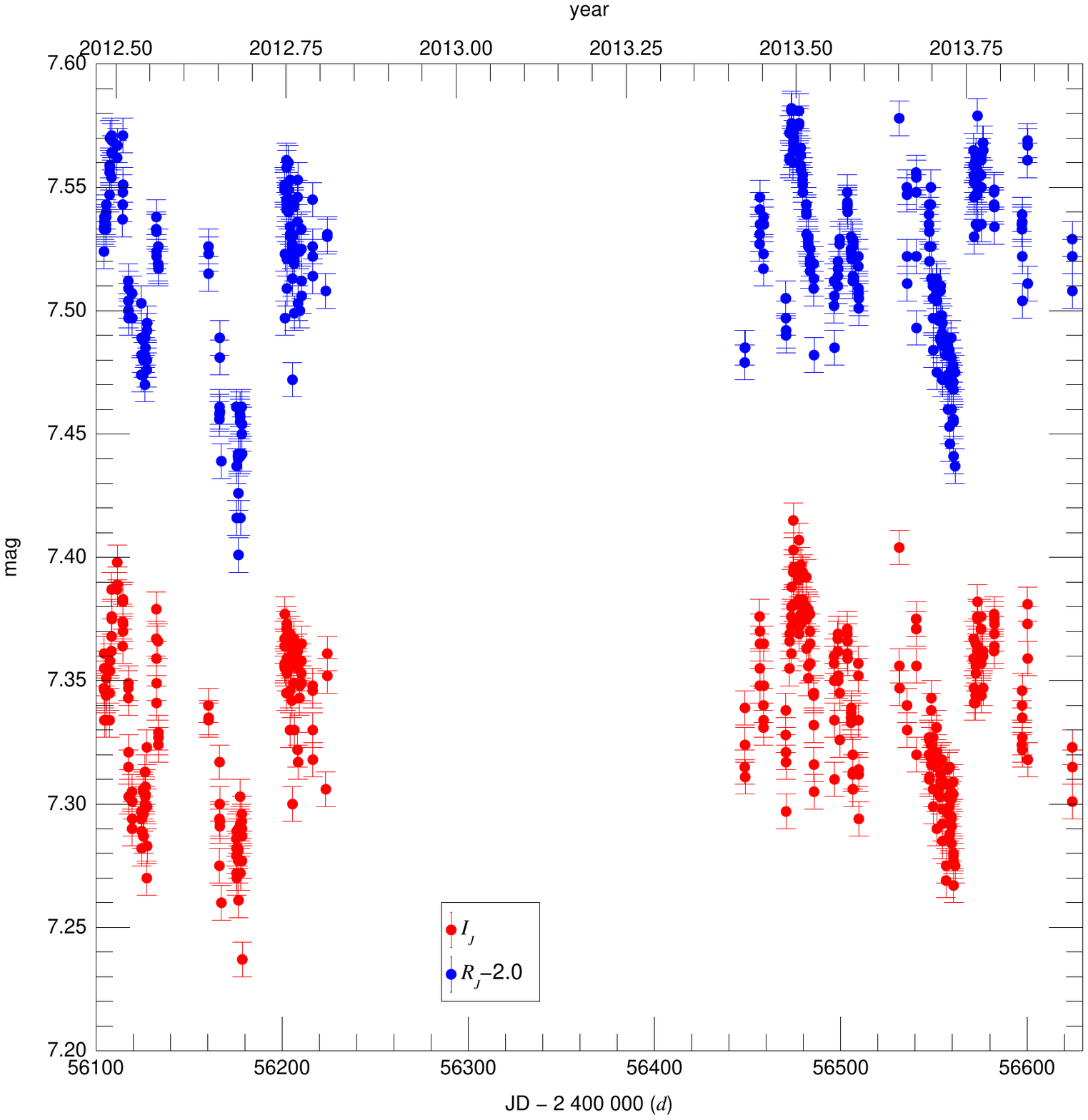} \
            \includegraphics*[width=0.49\linewidth, bb=28 28 566 566]{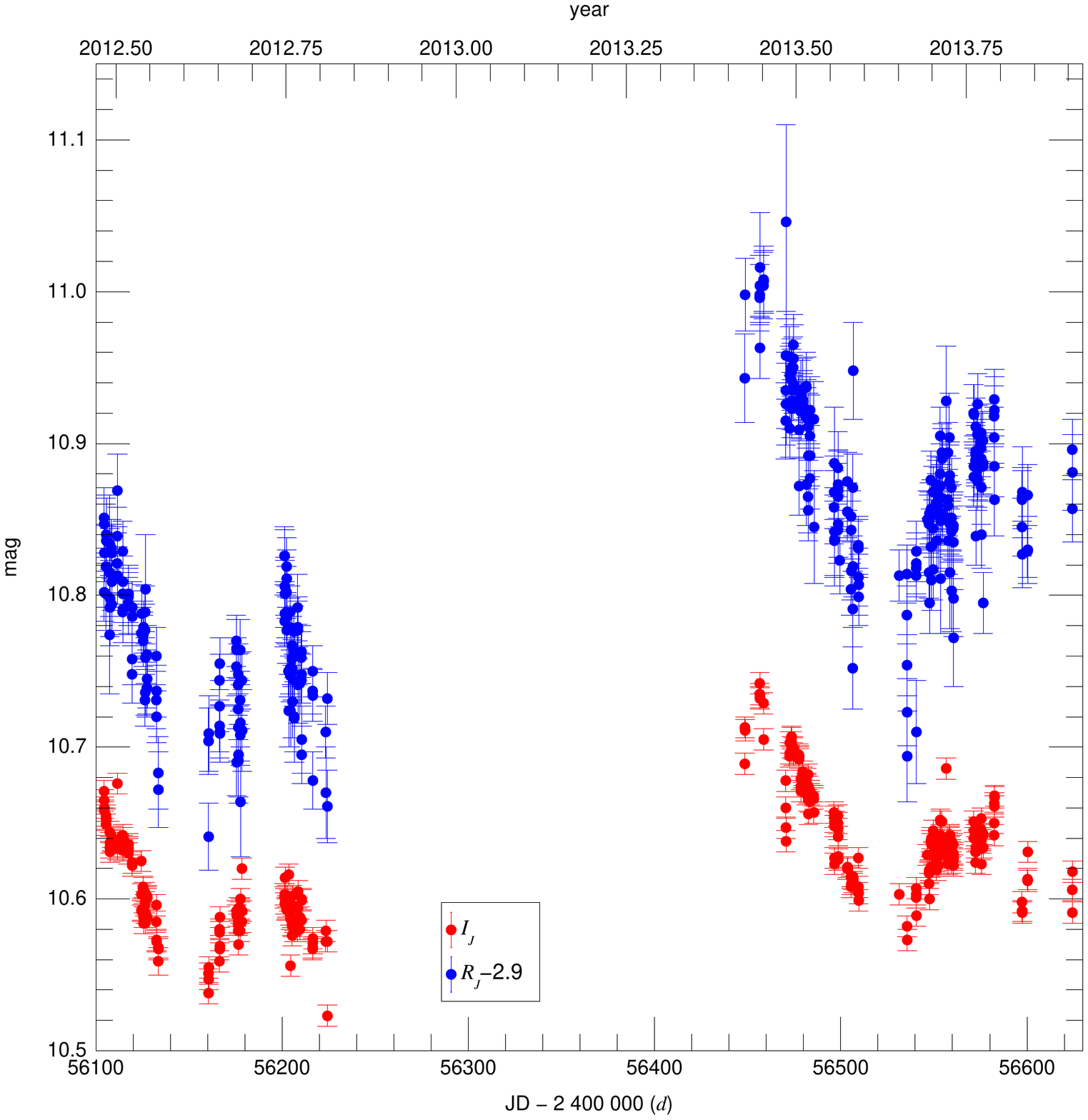}}
\caption{Light curves for two highly extinguished irregular variable stars, Cyg OB2-12 (left) and Cyg OB2-IRS 7 (right).}
\label{fig8}
\end{figure}

\section*{References}
\begin{itemize}
 \item \href{http://adsabs.harvard.edu/abs/1995ApJ...449..231C}{Cincotta, P.~M., M{\'e}ndez, M., \& N{\'u}{\~n}ez, J.~A. 1995, {\it ApJ} {\bf 449}, 231}.
 \item \href{http://adsabs.harvard.edu/abs/2011ApJS..194...27H}{Henderson, C.~B. et al. 2011, {\it ApJS} {\bf 194}, 27}.
 \item \href{http://adsabs.harvard.edu/abs/1986ApJ...302..757H}{Horne, J.~H. \& Baliunas, S.~L. 1986, {\it ApJ} {\bf 302}, 757}.
 \item \href{http://adsabs.harvard.edu/abs/2012ApJ...747...41K}{Kiminki, D.~C. et al. 2012, {\it ApJ} {\bf 747}, 41}.
 \item \href{http://adsabs.harvard.edu/abs/2013hsa7.conf..657M}{Ma{\'{\i}}z Apell{\'a}niz, J. 2013, {\it Highlights of Spanish Astrophysics VII}, 657}.
 \item \href{http://adsabs.harvard.edu/abs/2004ApJS..151..103M}{Ma{\'{\i}}z Apell{\'a}niz, J. et al. 2004, {\it ApJS} {\bf 151}, 103}.
 \item \href{http://adsabs.harvard.edu/abs/2011hsa6.conf..467M}{Ma{\'{\i}}z Apell{\'a}niz, J. et al. 2011, {\it Highlights of Spanish Astrophysics VI}, 467}.
 \item \href{http://adsabs.harvard.edu/abs/2014A&A...564A..63M}{Ma{\'{\i}}z Apell{\'a}niz, J. et al. 2014, {\it A\&A} {\bf 564}, A63}.
 \item \href{http://adsabs.harvard.edu/abs/1991AJ....101.1408M}{Massey, P. \& Thompson, A.~B. 1991, {\it AJ} {\bf 101}, 1408}.
 \item \href{http://adsabs.harvard.edu/abs/2013ATel.5571....1S}{Salas, J. et al. 2013, {\it ATel} 5571}.
 \item \href{http://adsabs.harvard.edu/abs/2008RMxAC..33...56S}{Sota, A. et al. 2008, {\it RvMxA\&A (conf. series)} {\bf 33}, 56}.
 \item \href{http://adsabs.harvard.edu/abs/2014MNRAS.438..639W}{Wright, N.~J. et al. 2014, {\it MNRAS} {\bf 438}, 639}.
\end{itemize}

\end{document}

%% file: mlaeff.tex
\def\hoy{\number\day \space de \space\ifcase\month\or
 Enero\or Febrero\or Marzo\or Abril\or Mayo\or Junio\or
 Julio\or Agosto\or Septiembre\or Octubre\or Noviembre\or Diciembre\fi
 \space de \number\year}
\def\ii/{\'{\i}}
\def\cion/{ci\'on}
\def\cao/{\c c\~ao}
\def\arcmin{\hbox{$^\prime$}}

\def\utw{\smash{\rlap{\lower5pt\hbox{$\sim$}}}}
\def\udtw{\smash{\rlap{\lower6pt\hbox{$\approx$}}}}

\def\tens#1{\ifmmode\mathchoice{\mbox{$\sf\displaystyle#1$}}
{\mbox{$\sf\textstyle#1$}}
{\mbox{$\sf\scriptstyle#1$}}
{\mbox{$\sf\scriptscriptstyle#1$}}\else
\hbox{$\sf\textstyle#1$}\fi}
\def\vec#1{\ifmmode\mathchoice{\mbox{\boldmath$\displaystyle#1$}}
{\mbox{\boldmath$\textstyle#1$}}
{\mbox{\boldmath$\scriptstyle#1$}}
{\mbox{\boldmath$\scriptscriptstyle#1$}}\else
\hbox{\boldmath$\textstyle#1$}\fi}
\def\bbbc{{\mathchoice {\setbox0=\hbox{$\displaystyle\rm C$}\hbox{\hbox
to0pt{\kern0.4\wd0\vrule height0.9\ht0\hss}\box0}}
{\setbox0=\hbox{$\textstyle\rm C$}\hbox{\hbox
to0pt{\kern0.4\wd0\vrule height0.9\ht0\hss}\box0}}
{\setbox0=\hbox{$\scriptstyle\rm C$}\hbox{\hbox
to0pt{\kern0.4\wd0\vrule height0.9\ht0\hss}\box0}}
{\setbox0=\hbox{$\scriptscriptstyle\rm C$}\hbox{\hbox
to0pt{\kern0.4\wd0\vrule height0.9\ht0\hss}\box0}}}}
\def\bbbq{{\mathchoice {\setbox0=\hbox{$\displaystyle\rm
Q$}\hbox{\raise
0.15\ht0\hbox to0pt{\kern0.4\wd0\vrule height0.8\ht0\hss}\box0}}
{\setbox0=\hbox{$\textstyle\rm Q$}\hbox{\raise
0.15\ht0\hbox to0pt{\kern0.4\wd0\vrule height0.8\ht0\hss}\box0}}
{\setbox0=\hbox{$\scriptstyle\rm Q$}\hbox{\raise
0.15\ht0\hbox to0pt{\kern0.4\wd0\vrule height0.7\ht0\hss}\box0}}
{\setbox0=\hbox{$\scriptscriptstyle\rm Q$}\hbox{\raise
0.15\ht0\hbox to0pt{\kern0.4\wd0\vrule height0.7\ht0\hss}\box0}}}}
\def\bbbt{{\mathchoice {\setbox0=\hbox{$\displaystyle\rm
T$}\hbox{\hbox to0pt{\kern0.3\wd0\vrule height0.9\ht0\hss}\box0}}
{\setbox0=\hbox{$\textstyle\rm T$}\hbox{\hbox
to0pt{\kern0.3\wd0\vrule height0.9\ht0\hss}\box0}}
{\setbox0=\hbox{$\scriptstyle\rm T$}\hbox{\hbox
to0pt{\kern0.3\wd0\vrule height0.9\ht0\hss}\box0}}
{\setbox0=\hbox{$\scriptscriptstyle\rm T$}\hbox{\hbox
to0pt{\kern0.3\wd0\vrule height0.9\ht0\hss}\box0}}}}
\def\bbbs{{\mathchoice
{\setbox0=\hbox{$\displaystyle     \rm S$}\hbox{\raise0.5\ht0\hbox
to0pt{\kern0.35\wd0\vrule height0.45\ht0\hss}\hbox
to0pt{\kern0.55\wd0\vrule height0.5\ht0\hss}\box0}}
{\setbox0=\hbox{$\textstyle        \rm S$}\hbox{\raise0.5\ht0\hbox
to0pt{\kern0.35\wd0\vrule height0.45\ht0\hss}\hbox
to0pt{\kern0.55\wd0\vrule height0.5\ht0\hss}\box0}}
{\setbox0=\hbox{$\scriptstyle      \rm S$}\hbox{\raise0.5\ht0\hbox
to0pt{\kern0.35\wd0\vrule height0.45\ht0\hss}\raise0.05\ht0\hbox
to0pt{\kern0.5\wd0\vrule height0.45\ht0\hss}\box0}}
{\setbox0=\hbox{$\scriptscriptstyle\rm S$}\hbox{\raise0.5\ht0\hbox
to0pt{\kern0.4\wd0\vrule height0.45\ht0\hss}\raise0.05\ht0\hbox
to0pt{\kern0.55\wd0\vrule height0.45\ht0\hss}\box0}}}}
\def\bbbz{{\mathchoice {\hbox{$\sf\textstyle Z\kern-0.4em Z$}}
{\hbox{$\sf\textstyle Z\kern-0.4em Z$}}
{\hbox{$\sf\scriptstyle Z\kern-0.3em Z$}}
{\hbox{$\sf\scriptscriptstyle Z\kern-0.2em Z$}}}}
\def\diameter{{\ifmmode\mathchoice
{\ooalign{\hfil\hbox{$\displaystyle/$}\hfil\crcr
{\hbox{$\displaystyle\mathchar"20D$}}}}
{\ooalign{\hfil\hbox{$\textstyle/$}\hfil\crcr
{\hbox{$\textstyle\mathchar"20D$}}}}
{\ooalign{\hfil\hbox{$\scriptstyle/$}\hfil\crcr
{\hbox{$\scriptstyle\mathchar"20D$}}}}
{\ooalign{\hfil\hbox{$\scriptscriptstyle/$}\hfil\crcr
{\hbox{$\scriptscriptstyle\mathchar"20D$}}}}
\else{\ooalign{\hfil/\hfil\crcr\mathhexbox20D}}%
\fi}}
\def\sq{\ifmmode\squareforqed\else{\unskip\nobreak\hfil
\penalty50\hskip1em\null\nobreak\hfil\squareforqed
\parfillskip=0pt\finalhyphendemerits=0\endgraf}\fi}
\def\squareforqed{\hbox{\rlap{$\sqcap$}$\sqcup$}}

%% file: eb.tex
\begin{tabular}{cclcr@{.}lr@{.}lr@{.}lc}
RA (J2000) & dec (J2000) & \multicolumn{1}{c}{Names} & Cand.? & \multicolumn{2}{c}{$R_J$} & \multicolumn{2}{c}{$I_J$} & \multicolumn{2}{c}{Period (d)} & $\Delta I_J$ \\
\hline
20:32:42.9 & 41:20:16 & ALS 21\,109, [MT91] 311             & --- & 13&0 & 12&0 & 6&2797(5)               & 0.14 \\
20:31:20.6 & 41:14:36 & 2MASS J20312066+4114363             & --- & 13&4 & 12&1 & 8&52651(5)              & 0.13 \\
20:31:23.4 & 41:19:25 & 2MASS J20312331+4119257             & --- & 16&3 & 14&8 & 7&9102(6)               & 0.42 \\
20:32:23.6 & 41:19:24 & [MT91] 242                          & --- & 14&8 & 14&1 & 5&6239(2)               & 0.52 \\
20:33:08.4 & 41:15:43 & [MT91] 849                          & --- & 16&2 & 15&1 & 2&8926(1)               & 0.35 \\
20:33:09.6 & 41:12:59 & Cyg OB2-22 C, [MT91] 421            & --- & 11&7 & 10&4 & 4&1621(1)               & 0.10 \\
20:33:30.5 & 41:20:17 & V2191 Cyg, [MT91] 554               & --- & 13&5 & 12&4 & 5&9504(3)               & 0.28 \\
20:31:27.8 & 41:29:17 & [MT91] 94                           & --- & 13&9 & 12&8 & 5&4669(5)               & 0.30 \\
20:31:41.6 & 41:28:21 & [MT91] 129                          & --- & 13&5 & 12&5 & 2&15175(3)              & 0.07 \\
20:33:10.5 & 41:22:22 & V2186 Cyg, [MT91] 429               & --- & 12&0 & 10&9 & 2&97864(6)              & 0.15 \\
20:31:37.5 & 41:13:20 & Cyg OB2-3 A,  BD +40 4212           & --- &  9&2 &  8&1 & 4&74565(5)              & 0.22 \\
20:31:59.0 & 41:07:31 & 2MASS J20315898+4107314             & --- & 13&0 & 11&6 & 2&53133(5)              & 0.22 \\
20:33:59.5 & 41:17:35 & Cyg OB2-27, [MT91] 696              & --- & 11&2 & 10&1 & 1&46917(2)              & 0.64 \\
20:34:06.0 & 41:08:08 & ALS 15\,146, [MT91] 720             & --- & 12&2 & 10&8 & 4&3619(1)               & 0.25 \\
20:34:41.4 & 41:07:45 & 2MASS J20344143+4107456             & --- & 14&8 & 13&3 & 2&89862(4)              & 0.26 \\
20:33:20.9 & 41:18:01 & V2189 Cyg, [MT91] 506               & --- & 15&5 & 14&4 & 1&31385(5)              & 0.30 \\
20:34:45.5 & 41:14:31 & 2MASS J20344555+4114314             & --- & 15&3 & 13&7 & 6&56500(5)              & 0.25 \\
20:32:38.3 & 41:28:56 & [MT91] 298                          &  Y  & 13&4 & 12&5 & \multicolumn{2}{c}{---} & 0.61 \\
20:32:34.1 & 41:22:55 & [MT91] 280                          &  Y  & 14&4 & 13&3 & \multicolumn{2}{c}{---} & 0.26 \\
20:32:11.3 & 41:25:04 & 2MASS J20321130+4125045             &  Y  & 16&5 & 15&5 & \multicolumn{2}{c}{---} & 0.60 \\
20:32:26.8 & 41:22:35 & [MT91] 254                          &  Y  & 15&8 & 14&8 & \multicolumn{2}{c}{---} & 1.42 \\
20:32:14.0 & 41:22:24 & 2MASS J20321399+4122240             &  Y  & 15&8 & 14&7 & \multicolumn{2}{c}{---} & 1.27 \\
20:31:57.0 & 41:12:33 & Tyc 3157-00779-1                    &  Y  & 12&2 & 11&7 & \multicolumn{2}{c}{---} & 0.32 \\
20:32:25.2 & 41:08:25 & [MT91] 245                          &  Y  & 13&8 & 13&3 & \multicolumn{2}{c}{---} & 0.26 \\
20:32:24.6 & 41:09:19 & 2MASS J20322463+4109202             &  Y  & 15&3 & 13&5 & \multicolumn{2}{c}{---} & 0.25 \\
20:33:18.5 & 41:24:38 & 2MASS J20331846+4124383             &  Y  & 11&8 & 11&2 & \multicolumn{2}{c}{---} & 0.10 \\
\hline
\end{tabular}

%% file: pul.tex
\begin{tabular}{cclcr@{.}lr@{.}lr@{.}lc}
RA (J2000) & dec (J2000) & \multicolumn{1}{c}{Names} & Cand.? & \multicolumn{2}{c}{$R_J$} & \multicolumn{2}{c}{$I_J$} & \multicolumn{2}{c}{Period (d)} & $\Delta I_J$ \\
\hline
20:33:04.3 & 41:24:39 & [MT91] 396                          & --- & 14&9 & 14&3 & 1&3823(3)               & 0.18 \\
20:31:51.3 & 41:23:23 & [MT91] 152                          & --- & 12&3 & 11&9 & 0&99838(4)              & 0.08 \\
20:31:23.6 & 41:29:49 & 2MASS J20312356+4129489             & --- & 15&7 & 14&7 & 0&9630(1)               & 0.17 \\
20:33:01.1 & 41:11:11 & [MT91] 382                          & --- & 14&2 & 13&7 & 0&93799(3)              & 0.26 \\
20:33:08.8 & 41:18:51 & Schulte 57                          & --- & 16&2 & 15&0 & 0&7275(1)               & 0.17 \\
20:33:30.8 & 41:15:22 & Cyg OB2-18, [MT91] 556              & --- &  9&7 &  8&3 & 1&1192(1)               & 0.07 \\
20:34:04.0 & 41:14:43 & 2MASS J20340404+4114430             & --- & 14&4 & 12&6 & 7&92(1)                 & 0.08 \\
20:33:23.0 & 41:12:22 & [MT91] 514                          & --- & 14&4 & 13&7 & 0&84634(1)              & 0.12 \\
20:33:18.3 & 41:17:39 & V2187 Cyg, [MT91] 487               & --- & 14&3 & 13&1 & 0&25385(2)              & 0.06 \\
20:33:13.3 & 41:13:28 & ALS 15\,148, [MT91] 448             & --- & 12&3 & 10&9 & 3&170(5)                & 0.03 \\
20:34:29.6 & 41:31:45 & ALS 15\,114, [MT91] 771             & --- & 10&7 &  9&4 & 1&4316(6)               & 0.02 \\
20:33:45.0 & 41:22:32 & [MT91] 626                          & --- & 12&2 & 11&7 & 1&11035(5)              & 0.06 \\
20:34:29.1 & 41:32:47 & 2MASS J20342909+4132476             & --- & 15&5 & 12&1 & 0&9833(2)               & 0.20 \\
20:33:43.0 & 41:30:00 & 2MASS J20334299+4130005             & --- & 14&2 & 13&0 & 0&95787(1)              & 0.17 \\
20:33:42.1 & 41:22:22 & [MT91] 617                          & --- & 14&0 & 13&3 & 1&6424(2)               & 0.08 \\
20:31:22.1 & 41:12:02 & Cyg OB2-A30                         & --- & 13&6 & 13&0 & 3&719(1)                & 0.04 \\
20:33:47.8 & 41:20:41 & Cyg OB2-26, [MT91] 642              &  Y  & 10&8 &  9&6 & \multicolumn{2}{c}{---} & 0.04 \\
20:33:18.0 & 41:18:31 & Cyg OB2-8 C, [MT91] 483             &  Y  &  9&3 &  8&4 & \multicolumn{2}{c}{---} & 0.03 \\
20:33:39.1 & 41:19:26 & Cyg OB2-19,  V1393 Cyg, [MT91] 601  &  Y  & 10&0 &  8&2 & \multicolumn{2}{c}{---} & 0.04 \\
20:35:02.5 & 41:21:27 & GSC 0316101176                      &  Y  & 13&9 & 13&4 & \multicolumn{2}{c}{---} & 0.23 \\
20:33:31.5 & 41:20:57 & [MT91] 916                          &  Y  & 15&4 & 14&1 & \multicolumn{2}{c}{---} & 0.33 \\
20:33:25.0 & 41:31:35 & Schulte 80                          &  Y  & 14&7 & 13&4 & \multicolumn{2}{c}{---} & 0.16 \\
20:33:49.8 & 41:23:58 & 2MASS J20334982+4123585             &  Y  & 15&1 & 13&8 & \multicolumn{2}{c}{---} & 0.35 \\
20:33:39.8 & 41:22:52 & [MT91] 605                          &  Y  & 10&9 &  9&9 & \multicolumn{2}{c}{---} & 0.06 \\
\hline
\end{tabular}

%% file: irr.tex
\begin{tabular}{cclcr@{.}lr@{.}lr@{.}lc}
RA (J2000) & dec (J2000) & \multicolumn{1}{c}{Names} & Cand.? & \multicolumn{2}{c}{$R_J$} & \multicolumn{2}{c}{$I_J$} & \multicolumn{2}{c}{Period (d)} & $\Delta I_J$ \\
\hline
20:32:03.7 & 41:25:10 & [MT91] 187                          & --- & 12&3 & 11&3 & 222&3(3)                & 0.07 \\
20:32:14.0 & 41:23:23 & 2MASS J20321405+4123237             & --- & 13&0 & 10&9 & 187&0(60)               & 0.06 \\
20:31:18.3 & 41:21:21 & ALS 15\,133, [MT91] 70              & --- & 11&7 & 10&3 & 192&0(40)               & 0.04 \\
20:31:39.4 & 41:21:38 & 2MASS J20313935+4121387             & --- & 14&5 & 12&3 & 126&0(10)               & 0.15 \\
20:32:30.8 & 41:10:00 & Cyg OB2-B12                         & --- & 14&1 & 12&4 & 52&8(2)                 & 0.15 \\
20:31:21.3 & 41:09:29 & 2MASS J20312131+4109286             & --- & 13&0 & 12&2 & 171&0(10)               & 0.08 \\
20:33:34.3 & 41:18:11 & [MT91] 575                          & --- & 12&1 & 10&8 & 209&0(90)               & 0.04 \\
20:33:31.7 & 41:18:53 & [MT91] 895                          & --- & 14&5 & 11&7 & 34&40(3)                & 0.09 \\
20:33:25.5 & 41:20:39 & [MT91] 911                          & --- & 15&0 & 12&2 & 59&5(6)                 & 0.14 \\
20:33:15.7 & 41:20:17 & Cyg OB2-23, [MT91] 470              & --- & 11&6 & 10&6 & 192&0(90)               & 0.04 \\
20:31:36.2 & 41:22:03 & Cyg OB2-A4                          & --- & 13&4 & 11&0 & 96&0(10)                & 0.13 \\
20:32:32.3 & 41:27:57 & 2MASS J20323232+4127571             & --- & 14&3 & 11&2 & 36&295(1)               & 0.27 \\
20:32:41.0 & 41:14:29 & Cyg OB2-12                          & --- &  9&5 &  7&3 & 54&0(1)                 & 0.18 \\
20:33:42.1 & 41:07:53 & [MT91] 615                          & --- & 11&0 & 10&5 & 309&0(90)               & 0.05 \\
20:33:39.6 & 41:10:18 & 2MASS J20333961+4110192             & --- & 14&4 & 11&0 & 73&5(2)                 & 0.33 \\
20:33:39.5 & 41:22:36 & Cyg OB2-IRS 7                       & --- & 13&8 & 10&6 & 109&0(10)               & 0.22 \\
20:31:43.1 & 41:06:56 & Tyc 3157-01040-1                    &  Y  & 10&5 & 10&1 & \multicolumn{2}{c}{---} & 0.18 \\
20:33:14.8 & 41:18:42 & Cyg OB2-8 B, [MT91] 462             &  Y  &  9&5 &  8&5 & \multicolumn{2}{c}{---} & 0.11 \\
\hline
\end{tabular}

%% file: Be.tex
\begin{tabular}{cclcr@{.}lr@{.}lr@{.}lc}
RA (J2000) & dec (J2000) & \multicolumn{1}{c}{Names} & Cand.? & \multicolumn{2}{c}{$R_J$} & \multicolumn{2}{c}{$I_J$} & \multicolumn{2}{c}{Period (d)} & $\Delta I_J$ \\
\hline
20:32:13.2 & 41:27:25 & Cyg OB2-4 B, [MT91] 213             & --- & 11&0 & 10&8 & \multicolumn{2}{c}{---} & 0.70 \\
20:33:18.5 & 41:15:35 & Schulte 64,  V2188 Cyg, [MT91] 488  & --- & 13&5 & 11&8 & \multicolumn{2}{c}{---} & 0.22 \\
20:34:43.6 & 41:29:04 & Schulte 30, [MT91] 793              & --- & 11&2 & 10&1 & \multicolumn{2}{c}{---} & 0.29 \\
\hline
\end{tabular}